\documentclass{ieeeaccess}
\usepackage{cite}
\usepackage{amsmath,amssymb,amsfonts}
\usepackage{algorithmic}
\usepackage{graphicx}
\usepackage{textcomp}

\usepackage{bm}
\usepackage{amsthm}
\usepackage{algorithm}
\def\BibTeX{{\rm B\kern-.05em{\sc i\kern-.025em b}\kern-.08em
    T\kern-.1667em\lower.7ex\hbox{E}\kern-.125emX}}
\begin{document}
\history{Received March 29, 2023, accepted April 17, 2023, date of publication xxxx 00, 0000, date of current version xxxx 00, 0000.}
\doi{}

\title{Resource Allocation in the RIS Assisted SCMA Cellular Network Coexisting with D2D Communications}
\author{\uppercase{Yukai Liu}\authorrefmark{1}, 
\uppercase{Wen Chen}\authorrefmark{1}, \IEEEmembership{Senior Member, IEEE}, \uppercase{Hongying Tang\authorrefmark{2}, and Kunlun Wang}\authorrefmark{3},
\IEEEmembership{Senior Member, IEEE}}
\address[1]{Shanghai Institute of Advanced Communications and Data Sciences, Department of Electronic Engineering, Shanghai Jiao Tong University, Shanghai 200240, China.}
\address[2]{Science and Technology on Microsystem Laboratory, Shanghai Institute of Microsystem and Information Technology, Chinese Academy of Sciences, Shanghai 200050, China.}
\address[3]{School of Communication and Electronic Engineering, East China Normal University, Shanghai 200241, China.}
\tfootnote{This work is supported by National key project 2020YFB1807700 and 2018YFB1801102, by Shanghai Kewei 20JC1416502 and 22JC1404000, and by NSFC 62071296.}

\markboth
{Y. Liu \headeretal: Resource Allocation in the RIS Assisted SCMA Cellular Network Coexisting with D2D Communications}
{Y. Liu \headeretal: Resource Allocation in the RIS Assisted SCMA Cellular Network Coexisting with D2D Communications}

\corresp{Corresponding author: Wen Chen (e-mail: wenchen@sjtu.edu.cn).}

\begin{abstract}
The cellular network coexisting with device-to-device (D2D) communications has been studied extensively. Reconfigurable intelligent surface (RIS) and non-orthogonal multiple access (NOMA) are promising technologies for the evolution of 5G, 6G and beyond. Besides, sparse code multiple access (SCMA) is considered suitable for next-generation wireless network in code-domain NOMA. In this paper, we consider the RIS-aided uplink SCMA cellular network simultaneously with D2D users. We formulate the optimization problem which aims to maximize the cellular sum-rate by jointly designing D2D users' resource block (RB) association, the transmitted power for both cellular users and D2D users, and the phase shifts at the RIS. The power limitation and users' communication requirements are considered. The problem is non-convex, and it is challenging to solve it directly. To handle this optimization problem, we propose an efficient iterative algorithm based on block coordinate descent (BCD) method. The original problem is decoupled into three subproblems to solve separately. Simulation results demonstrate that the proposed scheme can significantly improve the sum-rate performance over various schemes.
\end{abstract}


\titlepgskip=-15pt

\maketitle

\section{Introduction}
\label{Introduction}
\PARstart{T}{he} current wireless communication systems are aiming at massive connectivity and higher data rate. Some key technologies including device-to-device (D2D) communications and non-orthogonal multiple access techniques (NOMA) are being investigated [1]. NOMA can support massive user access via non-orthogonal resource block (RB) allocation [2], which is an overloaded system. Two main approaches of the power-domain NOMA and the code-domain NOMA are widely investigated. Among the currently proposed code domain NOMA, sparse code multiple access (SCMA) has been commonly considered as a promising method. SCMA is proposed by Nikopour and Baligh [3], in which, coded bits are directly mapped into the multi-dimensional complex lattice point (called codeword) and the codewords are designed to be sparse. The sparsity of the SCMA codewords enables massive connectivity and the use of suboptimal message passing algorithm (MPA) to detect multiple users. For general SCMA system, the users' SCMA codeword in each dimension will be modulated to an Orthogonal Frequency Division Multiple (OFDM) subcarrier before the air interface. The number of users is larger than the number of subcarriers, which lead to the non-orthogonal feature.

D2D communication is a promising technology for local communication, which allows nearby devices to communicate without base station (BS) or with limited BS involvement, and improves the link reliability, spectral efficiency and system capacity [4]. When D2D communications are considered in uplink SCMA cellular network, two main modes have been selected. In the simple mode, a proportion of available RBs are separately allocated to D2D users while the mutual interference between cellular users and D2D users can be reduced. In another mode, SCMA is only employed by cellular users and some subcarriers can be reused by D2D users' transmission. Liu $et$ $al.$ [5] have considered the analytical model for the SCMA enhanced cellular network coexisting with D2D. For large research work in these hybrid networks, scholars focus on the optimization problems while guaranteeing both users' communication requirements, but the mutual interference is always tough to handle.

Recently, reconfigurable intelligent surfaces (RIS) are becoming promising technologies towards many issues. RIS consists of many low-cost and passive elements. By dynamically changing the amplitudes and phase shifts on these elements, the propagation of incident signals can be modified [6]. Thus, the received signal can be enhanced after some optimization, which allows RIS to be deployed on multiple scenarios [7]. For example, some latest analysis works have proved the prominent features and advantages of RIS-aided internet of things (IoT) networks [8].

Many recent works have investigated the application of RIS in general NOMA systems [9]-[21]. Authors in [9] have compared downlink RIS-NOMA with spatial division multiple access (SDMA) and analyzed the influence of outage probability with different design schemes for amplitudes and phase shifts. Furthermore, the sum-rate optimization problems have been considered in [10]-[15]. Mu $et$ $al.$ [10] have focused on the optimal beamforming at both the BS and the RIS. For both the ideal RIS scenario and the non-ideal RIS scenario, they have proposed efficient algorithms considering the continuous phase shifts and discrete phase shifts. More subproblems like channel assignment, decoding order of NOMA users, power allocation, and reflection coefficients have been considered in [11], and the performance is compared with conventional NOMA system without RIS as well as RIS-OMA systems. However, [12] has formulated similar problems where an RIS is deployed for multiple cells NOMA networks instead of single cell. The uplink RIS-NOMA systems have been considered in [13] and [14], and [15] have proposed machine learning approaches for resource allocation problems.

Moreover, the transmission power allocation problems are also worth discussing [16]-[21]. Zhu $et$ $al.$ [16] have proposed a downlink multiple-input single-output (MISO) scheme for RIS-NOMA where the wireless channels can be effectively tuned. Both the beamforming vectors and the RIS phase shift matrix are optimized. The authors in [17] have pursued a theoretical performance comparison between time division multiple access (TDMA) and NOMA for uplink offloading, and proposed efficient algorithms for associated computation rate maximization problems. The two-cell structure has been considered in [18] to optimize both power allocation coefficients and phase shift matrix. Besides, the minimal power consumption problem and the maximal system energy efficiency problem have been considered in [19] and [20], respectively. The framework of RIS-enabled multi-group NOMA networks with coordinated multi-point (CoMP) reception and imperfect successive interference cancellation has been investigated [21].

\subsection{Related Works}
As SCMA is considered suitable for next-generation wireless network in code-domain NOMA, seldom works have investigated RIS-aided SCMA networks [22]-[24]. In these works, the uplink SCMA scenario has been considered, as users send data to the BS via the direct link (DL) and the RIS link. The diversity order has been analyzed in [22], as the pairwise error probability (PEP) is derived for both the condition of random discrete phase shifting and the condition of coherent phase shifting. However, the authors in [23] have utilized the modified MPA to decode the RIS-SCMA transmitted signal and proposed a low-complexity decoder. In order to improve the received signal-to-noise ratio (SNR) for discrete RIS phase shifts, some optimization problems have been formulated and solved with alternate optimization technique [24].

Some works have attempted to integrate RISs into general D2D communications [25]-[40]. Analytical results have been derived in [25] and [26], including the secrecy outage probability and the probability of non-zero secrecy capacity with the model of one cellular user and one D2D pair [25]. In [26], the authors have proposed novel closed-form expressions for outage probability under both overlay mode and underlay mode, where the Nakagami-m fading channel is considered. Furthermore, the system sum-rate maximization problems have been extensively considered in [27]-[31]. The RIS-aided scenario with one uplink cellular link and multiple D2D links has been utilized to jointly optimize the transmission power of all links as well as the discrete phase shifts of the surface [27], while [28] has considered multiple cellular users and added the resource reuse indicators for optimization. Ji $et$ $al.$ [29] have utilized decentralized double deep Q-network (D$^3$QN) framework and other machine learning approaches to solve the problem from a new perspective. Besides, the authors in [30] have presented a D2D-underlaid cellular system, where the multiple RISs are deployed at the cell boundary to improve propagation. RIS has also been introduced to joint processing coordinated multipoint (JP-CoMP) cellular networks with underlaying D2D communications to mitigate the interference [31].

Some RIS-aided D2D communication systems have focused on the improvement for communication quality of D2D users [32]-[35]. A RIS-aided D2D communication system which reuses the frequency and time resources assigned to a multiuser MISO downlink cellular transmission has been proposed in [33]. They have aimed to maximize the D2D ergodic weighted sum-rate subject to a given signal-to-interference-plus-noise ratio (SINR) target for each cellular user. However, [35] has investigated the RIS-aided system with all D2D links, and considered the Rician fading channels with imperfect hardware including both devices and RISs. Furthermore, the similar system model as in [28] has been considered to maximize the overall network's spectrum efficiency and energy efficiency [36]. The expression of secrecy rate for cellular users has been derived in [37] and the secrecy rate maximization problem has been formulated while satisfying the requirements of D2D communications. RIS can also help for D2D cooperative computing system to minimize the total computing delay [39].

\subsection{Motivation and Contribution}
To the best of the authors' knowledge, the SCMA cellular network coexisting with D2D has not been investigated under the RIS scenario yet [41], [43]. As the reason that the mutual interference between SCMA cellular users and D2D users is difficult to mitigate, the RIS can be more effective for enhancing the communication quality of SCMA cellular users. When the parameters of RIS are adjusted appropriately under SCMA systems, the interference from D2D user can be reduced in each RB, which improves the transmission rate of cellular user. This paper considers a RIS-aided cellular uplink network simultaneously with D2D communications. The complex mode is utilized as D2D pairs will reuse the RBs of SCMA cellular users.

In this paper, our main contribution contains the problem formulation for cellular users' sum-rate maximization optimization, as well as the proposed iterative algorithm for solving that problem. Specifically, we propose an uplink RIS-aided hybrid network including cellular users and D2D users. We derive and analyze the received signal for both cellular transmission and D2D transmission according to the proposed network model. Then, we focus on the optimization of various resources to maximize cellular sum-rate, where users' transmitted power, RB resource allocation for D2D, and reconfigurable RIS phase shifts are included. For the problem formulation, we take power limitation and D2D's communication requirements into consideration. We analyze that the problem is mixed-integer non-linear programming (MINLP) problem which is NP-hard. Then, our proposed algorithm is based on block coordinate descent (BCD) method to find the suboptimal solution. The original problem is decoupled into three subproblems and we propose adequate analysis for solving these subproblems. Besides, the convergence as well as computational complexity are analyzed from a theoretical perspective to prove that our overall algorithm is converged after iterations. Furthermore, the benefits of the proposed algorithm are conducted by simulation part. We compare the performance from different aspects to the hybrid network without RIS in order to show the improvement of this RIS-aided network effectively. The proposed algorithm is also appropriate and significant after the comparison with some benchmark methods.

\subsection{Organization and Notation}
The rest of the paper is organized as follows. The system model and the optimization problem formulation are presented in Section II. In Section III, the effective algorithms are proposed to solve the holistic resource allocation problem. Section IV devotes to the numerical results, and the paper finally concludes in Section V.

Throughout this paper, the following notations will be used [41]. The bold lowercase letter $\bm{x}$ denotes a column vector, and a matrix is represented by a bold uppercase letter $\bm{X}$. $\bm{I}$ denotes the identity matrix. The superscript $(\cdot)^{T}$ denotes matrix transpose and $(\cdot)^{H}$ denotes conjugate transpose matrix. $diag(\bm{x})$ denotes a diagonal matrix with the diagonal entries being vector $\bm{x}$. Besides, $rank(\bm{X})$ and $Tr(\bm{X})$ denote the rank and trace of the matrix $\bm{X}$, respectively. For a set $\textit{A}$, $|\textit{A}|$ denotes the number of elements, and $\textit{A}\setminus n$ denotes the set $\textit{A}$ in which the element $n$ is excluded.

\section{System Model And Problem Formulation}
\label{System}
In this section, we introduce the hybrid RIS-aided network with cellular users and D2D users. The specific analysis for both cellular transmission and D2D transmission are presented. After that, we focus on the cellular sum-rate maximization problem to optimize a series of resource variables.

\subsection{RIS-aided Hybrid Network}
\label{hybrid}
\begin{figure}
	\centering  
	\includegraphics[width=0.7\linewidth]{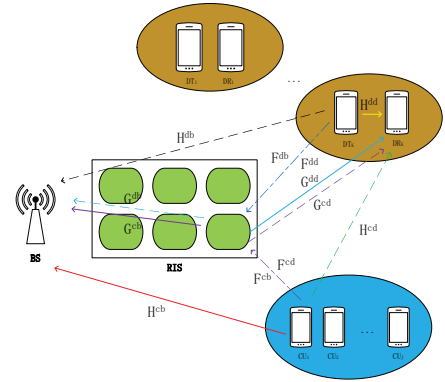}  
	\caption{The RIS-aided cellular and D2D hybrid network.}  
	\label{fig11}
\end{figure}

Fig.~\ref{fig11} shows an uplink RIS-aided hybrid network example of cellular users coexisting with two D2D pairs. In our proposed network, we consider $J$ single-antenna cellular users and $J_{D}$ single-antenna D2D pairs. The $i$th cellular user is denoted as CU$_{i}$, and the D2D transmitter as well as the D2D receiver of the $i$th D2D pair are denoted as DT$_{i}$ and DR$_{i}$, respectively. Cellular users are allocated with SCMA codewords spread over $K$ OFDM tones and transmitted to the air interface. While for each D2D pair, one of the same OFDM subcarriers will be occupied so that $J_{D} \leq K$. Besides, the RIS with $M$ reflecting elements is adopted to improve uplink cellular communications. Let $\bm{\Theta}=diag(e^{j\theta_{1}},e^{j\theta_{2}},...,e^{j\theta_{M}})$ denotes the phase shift matrix of the RIS, where $\theta_{m}$ is the phase shift coefficient for $m$th element with $\theta_{m} \in (0,2\pi]$. All elements remain the amplitude of incident signal unchanged.

\subsection{Cellular Transmission}
\label{ybs}
The SCMA codeword vector $\bm{x}_{j}^{c}$ of user $CU_{j}$ is $\bm{x}_{j}^{c}=(x_{j1}^{c},x_{j2}^{c},...,x_{jK}^{c})^{T}$. Besides, the D2D transmitted codeword for user $DT_{\ell}$ is denoted as $x_{\ell}^{d}$. The power of signal $x_{jk}^{c}$ and $x_{\ell}^{d}$ are denoted as $p_{jk}^{c}$ and $p_{\ell}^{d}$, respectively. For the SCMA enhanced cellular uplink transmission, the received signal $\bm{y}^{c}$ at BS should be a $K$-dimensional vector, and for $k=1,2,...,K$ we have
\begin{equation}
\begin{aligned}
&y_{k}^{c}=\sum_{j\in\xi_{k}}(h_{jk}^{cb}+(\bm{g}_{jk}^{cb})^{H}\bm{\Theta}\bm{f}_{jk}^{cb})\sqrt{p_{jk}^{c}}x_{jk}^{c}\\
&+\sum_{\ell=1}^{J_{D}}a_{\ell k}(h_{\ell k}^{db}+(\bm{g}_{\ell k}^{db})^{H}\bm{\Theta}\bm{f}_{\ell k}^{db})\sqrt{p_{\ell}^{d}}x_{\ell}^{d}+n_{k}.
\label{eq1}
\end{aligned}
\end{equation}
In \eqref{eq1}, $h_{jk}^{cb}$ denotes the channel for $CU_{j}$-to-BS over the $k$th RB, $\bm{f}_{jk}^{cb}$ and $\bm{g}_{jk}^{cb}$ are $M\times1$ complex-valued vectors which denote the channels for $CU_{j}$-to-RIS and RIS-to-BS over the $k$th RB, respectively. Thus, the first part is the desired signal with cellular power. The second part of \eqref{eq1} denotes the interference of D2D signal, where $h_{\ell k}^{db}$ is the channel for $DT_{\ell}$-to-BS over the $k$th RB, $\bm{f}_{\ell k}^{db}$ and $\bm{g}_{\ell k}^{db}$ are the channels for $DT_{\ell}$-to-RIS and RIS-to-BS over the $k$th RB, respectively. The $a_{\ell k}$ is an indicator to show whether the $\ell$th D2D pair occupy the $k$th RB. In the last part, $n_{k}\sim\bm{\textit{CN}}(0,N_{0})$ is the additive white Gaussian noise (AWGN) over the $k$th RB. We can notice that the desired signal at BS in each RB do not include all cellular users, but choose particular users according to SCMA structure [3].

For simplify, we define that for $\forall j=1,2,...,J$, $\forall k=1,2,...,K$, and $\forall \ell=1,2,...,J_{D}$,
\begin{equation}
h_{jk}^{cb}+(\bm{g}_{jk}^{cb})^{H}\bm{\Theta}\bm{f}_{jk}^{cb}\triangleq h_{jk}^{CB}(\bm{\Theta}),
\label{eq2}
\end{equation}
and
\begin{equation}
h_{\ell k}^{db}+(\bm{g}_{\ell k}^{db})^{H}\bm{\Theta}\bm{f}_{\ell k}^{db}\triangleq h_{\ell k}^{DB}(\bm{\Theta}).
\label{eq3}
\end{equation}
Most SCMA system models utilize optimal codebook design principles to obtain the diagonal covariance matrix [42]. Accordingly, the achievable capacity for all cellular users is [43]
\begin{equation}
R^{c}=\sum_{k=1}^{K}\log\left(1+\frac{\sum_{j\in\xi_{k}}|h_{jk}^{CB}(\bm{\Theta})|^{2}p_{jk}^{c}}{\sum_{\ell=1}^{J_{D}}a_{\ell k}|h_{\ell k}^{DB}(\bm{\Theta})|^{2}p_{\ell}^{d}+N_{0}}\right),
\label{eq4}
\end{equation}
where $a_{\ell k}=1$ means the $\ell$th D2D link occupies the $k$th RB, and $a_{\ell k}=0$ otherwise. For decoding the codewords of the $CU_{j}$ at the BS, we define the signal-to-interference-plus-noise ratio (SINR) in subcarrier $k\in\zeta_{j}$ according to SCMA structure as
\begin{equation}
\gamma_{jk}^{c}=\frac{|h_{jk}^{CB}(\bm{\Theta})|^{2}p_{jk}^{c}}{\sum_{\ell=1}^{J_{D}}a_{\ell k}|h_{\ell k}^{DB}(\bm{\Theta})|^{2}p_{\ell}^{d}+N_{0}}.
\label{eq5}
\end{equation}

\subsection{D2D Transmission}
\label{ydr}
For D2D communications, each D2D pair is allocated with one of the $K$ OFDM tones. For example, we assume that the $\ell$th D2D pair selects the $\ell$th OFDM tone for $\ell=1,2,...,J_{D}$. Notice that although we utilize RIS for cellular communications, these surfaces also make impact on D2D transmission. Thus, the received signal at the $DR_{\ell}$ is shown as
\begin{equation}
\begin{aligned}
&y_{\ell}^{d}=(h_{\ell\ell}^{dd}+(\bm{g}_{\ell\ell}^{dd})^{H}\bm{\Theta}\bm{f}_{\ell\ell}^{dd})\sqrt{p_{\ell}^{d}}x_{\ell}^{d}\\
&+\sum_{j\in\xi_{\ell}}(h_{j\ell}^{cd}+(\bm{g}_{j\ell}^{cd})^{H}\bm{\Theta}\bm{f}_{j\ell}^{cd})\sqrt{p_{j\ell}^{c}}x_{j\ell}^{c}+n_{\ell},
\end{aligned}
\label{eq6}
\end{equation}
where $h_{\ell\ell}^{dd}$, $\bm{g}_{\ell\ell}^{dd}$ and $\bm{f}_{\ell\ell}^{dd}$ are the channels for $DT_{\ell}$-to-$DR_{\ell}$, $DT_{\ell}$-to-RIS, and RIS-to-$DR_{\ell}$ over the $\ell$th RB, respectively. Besides, $h_{j\ell}^{cd}$, $\bm{g}_{j\ell}^{cd}$ and $\bm{f}_{j\ell}^{cd}$ are the channels for $CU_{j}$-to-$DR_{\ell}$, $CU_{j}$-to-RIS and RIS-to-$DR_{\ell}$ over the $\ell$th RB, respectively. $n_{\ell}$ is still the AWGN over the $\ell$th RB. The signal in \eqref{eq6} also includes the desired D2D signal and the interference of both cellular users and noise.

Generally, we define the SINR for the $\ell$th D2D receiver as
\begin{equation}
\begin{aligned}
&\gamma_{\ell}^{d}\\
&=\frac{\sum_{k=1}^{K}a_{\ell k}|h_{\ell k}^{dd}+(\bm{g}_{\ell k}^{dd})^{H}\bm{\Theta}\bm{f}_{\ell k}^{dd}|^{2}p_{\ell}^{d}}{\sum_{k=1}^{K}a_{\ell k}(\sum_{j\in\xi_{k}}|h_{jk}^{cd}+(\bm{g}_{jk}^{cd})^{H}\bm{\Theta}\bm{f}_{jk}^{cd}|^{2}p_{jk}^{c})+N_{0}}\\
&\triangleq \frac{\sum_{k=1}^{K}a_{\ell k}|h_{\ell k}^{DD}(\bm{\Theta})|^{2}p_{\ell}^{d}}{\sum_{k=1}^{K}a_{\ell k}(\sum_{j\in\xi_{k}}|h_{jk}^{CD}(\bm{\Theta})|^{2}p_{jk}^{c})+N_{0}}.
\end{aligned}
\label{eq7}
\end{equation}

\subsection{Problem Formulation}
In our hybrid network with cellular and D2D users, we aim to introduce RIS to enhance SCMA cellular communication qualities. Thus, the target of our problem is to maximize cellular users' sum-rate by jointly optimizing both users' transmitted power, RIS phase shift matrix, and the RB allocation of D2D pairs. The optimization problem is formulated as
\begin{equation}
\begin{aligned}
\bm{P1:} \quad & \max_{\bm{P}^{c},\bm{p}^{d},\bm{\Theta},\bm{A}} R^{c}\\
s.t.\quad & C1:\gamma_{\ell}^{d}\geq\gamma_{0}^{d},\quad \ell=1,\dots,J_{D},\\
\quad& C2:\sum_{\ell=1}^{J_{D}}a_{\ell k}\leq 1,\quad k=1,\dots,K,\\
\quad& C3:\sum_{k=1}^{K}a_{\ell k}= 1,\quad \ell=1,\dots,J_{D},\\
\quad& C4:0\leq p_{j}^{c}=\sum_{k\in\zeta_{j}}p_{jk}^{c}\leq P_{0}^{c},\quad j=1,\dots,J,\\
\quad& C5:0\leq p_{\ell}^{d}\leq P_{0}^{d},\quad \ell=1,\dots,J_{D},\\
\quad& C6:0<\theta_{m}\leq 2\pi,\quad m=1,\dots,M,
\end{aligned}
\label{eq8}
\end{equation}
where constraint $C1$ denotes the minimum QoS requirements for D2D communications, constraints $C2$ and $C3$ show that each D2D pair can only be allocated with one RB and each RB cannot be reused. Besides, $C4$ and $C5$ denote that cellular users' maximum power and D2D users' maximum power are upper bounded by $P_{0}^{c}$ and $P_{0}^{d}$, respectively. $C6$ is the practical constraint for RIS design.

We can notice that since the $a_{\ell k}$ is integer variable, and the objective function is non-convex for either variables $\bm{P}^{c}$, $\bm{p}^{d}$ or $\bm{\Theta}$, problem $\bm{P1}$ is MINLP problem, which is NP-hard. Besides, the exhaustive search is infeasible as the computational complexity is extremely high. To the best of the authors' knowledge, there is no perfect methods to find the global optimal solution for this kind of problems. In the next section, we aim to propose effective algorithms based on BCD for the feasible suboptimal solution.

\section{Proposed Algorithm for Optimization Problem}
\label{Method}
Notice that our optimization problem focus on coupled variables, the first step is to decouple $\bm{P1}$ into three parts to get that tractable. All optimization variables are divided into three categories: $\bm{\mathcal{P}}$, $\bm{A}$, and $\bm{\Theta}$, where the $\bm{\mathcal{P}}$ denotes the variable set for power of both cellular users and D2D users. In each part, we transform the problem $\bm{P1}$ into a subproblem where only one class of variables are needed to optimize. Thus, we propose specific analysis for solving these subproblems.

\subsection{Subproblem of Power Allocation}
\label{pforp1}
For this part, we consider the cellular sum-rate maximization problem with given $\bm{A}$ and $\bm{\Theta}$. Accordingly, the subproblem $\bm{P2}$ is formulated as
\begin{equation}
\begin{aligned}
&\bm{P2:}\\
& \max_{\bm{\mathcal{P}}}\quad \sum_{k=1}^{K}\log\left(1+\frac{\sum_{j\in\xi_{k}}|h_{jk}^{CB}(\bm{\Theta})|^{2}p_{jk}^{c}}{\sum_{\ell=1}^{J_{D}}a_{\ell k}|h_{\ell k}^{DB}(\bm{\Theta})|^{2}p_{\ell}^{d}+N_{0}}\right)\\
&s.t.\quad C1:\gamma_{\ell}^{d}\geq\gamma_{0}^{d},\quad \ell=1,\dots,J_{D},\\
&\quad C4:0\leq p_{j}^{c}=\sum_{k\in\zeta_{j}}p_{jk}^{c}\leq P_{0}^{c},\quad j=1,\dots,J,\\
&\quad C5:0\leq p_{\ell}^{d}\leq P_{0}^{d},\quad \ell=1,\dots,J_{D},\\
\end{aligned}
\label{eq9}
\end{equation}
where the constraints are reduced as some variables are fixed. It is not hard to see that $\bm{P2}$ is still non-convex to $\bm{\mathcal{P}}$ in objective function and constraint $C1$, so we introduce a new variable set $\bm{\mathcal{Q}}=\{Q_{1},...,Q_{K}\}$ to handle the complicated terms $\Gamma_{K}(\bm{\mathcal{P}})$. Where
\begin{equation}
\Gamma_{k}(\bm{\mathcal{P}})\triangleq\frac{\sum_{j\in\xi_{k}}|h_{jk}^{CB}(\bm{\Theta})|^{2}p_{jk}^{c}}{\sum_{\ell=1}^{J_{D}}a_{\ell k}|h_{\ell k}^{DB}(\bm{\Theta})|^{2}p_{\ell}^{d}+N_{0}}.
\label{eq10}
\end{equation}

Notice that in constraint $C1$, when $\bm{A}$ is fixed, the form of $\gamma_{\ell}^{d}$ can be rewritten as
\begin{equation}
\gamma_{\ell}^{d}=\frac{|h_{\ell k^{*}}^{DD}(\bm{\Theta})|^{2}p_{\ell}^{d}}{\sum_{j\in\xi_{k^{*}}}|h_{jk^{*}}^{CD}(\bm{\Theta})|^{2}p_{jk^{*}}^{c}+N_{0}},
\label{eq11}
\end{equation}
where we assume that $a_{\ell k^{*}}=1$. Thus, all inequalities in $C1$ can be changed to the format as $u_{\ell}p_{\ell}^{d}+\bm{v}_{\ell}^{T}\bm{p}^{c}+w_{\ell}\geq 0$, which is always convex to all power variables.	

Based on the analysis of \eqref{eq10} and \eqref{eq11}, we further formulate the problem $\bm{P3}$ as
\begin{equation}
\begin{aligned}
\bm{P3:}\quad &\max_{\bm{\mathcal{P}},\bm{\mathcal{Q}}}\quad \sum_{k=1}^{K}\log(1+Q_{k})\\
s.t.\quad &C1, C4, C5,\\
\quad &C7:\frac{\sum_{j\in\xi_{k}}|h_{jk}^{CB}(\bm{\Theta})|^{2}p_{jk}^{c}}{\sum_{\ell=1}^{J_{D}}a_{\ell k}|h_{\ell k}^{DB}(\bm{\Theta})|^{2}p_{\ell}^{d}+N_{0}}\geq Q_{k}.\\
\end{aligned}
\label{eq12}
\end{equation}
\\
	\label{prop1}
	$\textit{Proposition}$ 1: Problem $\bm{P3}$ is equivalent to $\bm{P2}$ for the optimal solution $\bm{\mathcal{P}}^{*}$.

Proof: When we rewrite the problem $\bm{P2}$ as $\bm{P3}$ , the $\Gamma_{k}(\bm{\mathcal{P}})$ is replaced by $Q_{k}$ and we should have the equality constraint. However, we relax that into the inequality constraint as $C7$ to optimize $Q_{k}$.
Notice that $\log(1+Q_{k})$ is monotonic increasing over variable $Q_{k}$, when the objective function in $\bm{P3}$ achieves its optimal value, the equality in $C7$ must hold. Thus, the problem $\bm{P2}$ and $\bm{P3}$ have same optimal solution $\bm{\mathcal{P}}^{*}$.

The objective function is convex to all variables in $\bm{P3}$, while the added constraint $C7$ belongs to non-convex condition. Specifically, we rewrite $C7$ as
\begin{equation}
\sum_{j\in\xi_{k}}|h_{jk}^{CB}(\bm{\Theta})|^{2}p_{jk}^{c}\geq\sum_{\ell=1}^{J_{D}}\tau_{\ell k}p_{\ell}^{d}Q_{k}+N_{0}Q_{k},
\label{eq13}
\end{equation}
where $\tau_{\ell k}\triangleq a_{\ell k}|h_{\ell k}^{DB}(\bm{\Theta})|^{2}$, and the term with $p_{\ell}^{d}Q_{k}$ is quasi-concave. In the following, we use convex upper bound (CUB) approximation to solve that non-convex part as shown in Lemma 1.
\\
	\label{ll1}
	$\textit{Lemma}$ 1: Define two functions $g(x,y)\triangleq xy$ and $f(x,y)\triangleq \frac{\alpha}{2}x^{2}+\frac{1}{2\alpha}y^{2}$ with $\alpha>0$, we always have $f(x,y)\geq g(x,y)$. Besides, when $\alpha=\frac{y}{x}$, we have $f(x,y)=g(x,y)$ and $\bigtriangledown f(x,y)=\bigtriangledown g(x,y)$, where $\bigtriangledown f(x,y)$ is the gradient of function $f(x,y)$.

This lemma is easy to prove as $f(x,y)$ is the CUB function of $g(x,y)$ [11], [12], and we aim to find the CUB for term with $p_{\ell}^{d}Q_{k}$. The basic idea is that we need the iterative process to improve the accuracy of CUB approximation. We set the initial value for $p_{\ell}^{d(0)}$ and $Q_{k}^{(0)}$, and let $\alpha_{\ell k}^{(1)}=\frac{p_{\ell}^{d(0)}}{Q_{k}^{(0)}}$, Then we have
\begin{equation}
p_{\ell}^{d}Q_{k}=\frac{1}{2\alpha_{\ell k}^{(1)}}(p_{\ell}^{d})^{2}+\frac{\alpha_{\ell k}^{(1)}}{2}(Q_{k})^{2}.
\label{eq14}
\end{equation}
Now we focus on the problem with changed constraint, which is shown as
\begin{equation}
\begin{aligned}
&\bm{P4:}\quad \max_{\bm{\mathcal{P}},\bm{\mathcal{Q}}}\quad \sum_{k=1}^{K}\log(1+Q_{k})\\
&s.t.\quad C1, C4, C5,\\
&C8:\sum_{j\in\xi_{k}}|h_{jk}^{CB}(\bm{\Theta})|^{2}p_{jk}^{c}\geq N_{0}Q_{k}+\\
&\quad\sum_{\ell=1}^{J_{D}}\tau_{\ell k}\left(\frac{1}{2\alpha_{\ell k}}(p_{\ell}^{d})^{2}+\frac{\alpha_{\ell k}}{2}(Q_{k})^{2}\right). \\
\end{aligned}
\label{eq15}
\end{equation}
$\bm{P4}$ is a convex problem which can be solved efficiently by standard algorithms or software, such as the SeDuMi solver in Matlab CVX. It is proved that after iterations the optimal solution of $\bm{P4}$ will converge to the solution of $\bm{P3}$ [11], and finally we can solve the power allocation problem in $\bm{P2}$.

Generally, the random initial values for $p_{\ell}^{d(0)}$ and $Q_{k}^{(0)}$ are infeasible for the iterative algorithm. We need another method to find the feasible initial points, and we still utilize optimization theory for help. $\bm{P5}$ is formulated to find the feasible $p_{\ell}^{d(0)}$ and $Q_{k}^{(0)}$.
\begin{equation}
\begin{aligned}
&\bm{P5:}\quad \min_{\bm{\mathcal{P}},\bm{\mathcal{Q}},z}\quad z\\
&s.t.\quad C9:\gamma_{\ell}^{d}+z\geq\gamma_{0}^{d},\\
&C10:-z\leq p_{j}^{c}=\sum_{k\in\zeta_{j}}p_{jk}^{c}\leq P_{0}^{c}+z, \\
&C11:-z\leq p_{\ell}^{d}\leq P_{0}^{d}+z,\quad\\
&C12:\sum_{j\in\xi_{k}}|h_{jk}^{CB}(\bm{\Theta})|^{2}p_{jk}^{c}+z\geq N_{0}Q_{k}+\\
&\quad \sum_{\ell=1}^{J_{D}}\tau_{\ell k}\left(\frac{1}{2\alpha_{\ell k}}(p_{\ell}^{d})^{2}+\frac{\alpha_{\ell k}}{2}(Q_{k})^{2}\right),\\
&C13:z\geq 0.
\end{aligned}
\label{eq16}
\end{equation}
The indicator $z$ denotes how long we could find the feasible points to some extent, and when $z=0$ after some iterations, the optimal solutions of $\bm{P5}$ is the appropriate initial values. The iterative algorithm for power allocation subproblem is summarized in Algorithm~\ref{algorithm1}.
\begin{algorithm}[h]
	\caption{Iterative power allocation algorithm.}
	\begin{algorithmic}[1]
		\STATE Initialize random value for $\bm{\mathcal{P}}$, $\bm{\mathcal{Q}}$ and set the maximum iterative number $T_{1}$ and $T_{2}$.
		\STATE Set $I_{1}=1$ and calculate $\alpha_{\ell k}=\frac{p_{\ell}^{d}}{Q_{k}}$.
		\WHILE {$I_{1}\leq T_{1}$}
		\STATE Solve problem $\bm{P5}$ to find optimal $p_{\ell}^{d(I_{1})}$ and $Q_{k}^{(I_{1})}$, and calculate $\alpha_{\ell k}^{(I_{1}+1)}=\frac{p_{\ell}^{d(I_{1})}}{Q_{k}^{(I_{1})}}$.
		\STATE Update $\bm{P5}$ with new $\alpha_{\ell k}^{(I_{1}+1)}$.
		\STATE $I_{1}=I_{1}+1$.
		\ENDWHILE
		\STATE Get the feasible initial values of $p_{\ell}^{d(0)}$ and $Q_{k}^{(0)}$.
		\STATE Set $I_{2}=1$ and calculate $\alpha_{\ell k}^{(1)}=\frac{p_{\ell}^{d(0)}}{Q_{k}^{(0)}}$.
		\WHILE {$I_{2}\leq T_{2}$}
		\STATE Solve the optimization problem $\bm{P4}$ to find the optimal solution $\bm{\mathcal{P}}^{*(I_{2})}$, $\bm{\mathcal{Q}}^{*(I_{2})}$.
		\STATE Calculate $\alpha_{\ell k}^{(I_{2}+1)}=\frac{p_{\ell}^{d(I_{2})}}{Q_{k}^{(I_{2})}}$ and update problem $\bm{P4}$ with $\alpha_{\ell k}^{(I_{2}+1)}$.
		\STATE $I_{2}=I_{2}+1$.
		\ENDWHILE
		\RETURN The optimal solution $\bm{\mathcal{P}}^{*}$ for $\bm{P2}$.
	\end{algorithmic}
	\label{algorithm1}
\end{algorithm}

\subsection{Subproblem of RB Allocation}
\label{aforp1}
In this part, the main variables are the indicators for D2D users' RB allocation. When the variables $\bm{\mathcal{P}}$ and $\bm{\Theta}$ are given, the subproblem $\bm{P6}$ is shown as
\begin{equation}
\begin{aligned}
&\bm{P6:}\\
&\max_{\bm{A}}\quad \sum_{k=1}^{K}\log\left(1+\frac{\sum_{j\in\xi_{k}}|h_{jk}^{CB}(\bm{\Theta})|^{2}p_{jk}^{c}}{\sum_{\ell=1}^{J_{D}}a_{\ell k}|h_{\ell k}^{DB}(\bm{\Theta})|^{2}p_{\ell}^{d}+N_{0}}\right)\\
&s.t.\quad C1:\gamma_{\ell}^{d}\geq\gamma_{0}^{d},\quad \ell=1,\dots,J_{D},\\
&\quad C2:\sum_{\ell=1}^{J_{D}}a_{\ell k}\leq 1,\quad k=1,\dots,K,\\
&\quad C3:\sum_{k=1}^{K}a_{\ell k}= 1,\quad \ell=1,\dots,J_{D}.\\
\end{aligned}
\label{eq17}
\end{equation}
Notice that the variable $\bm{A}$ is a $J_{D}\times K$ matrix with all $\{0,1\}$ elements, it is difficult to solve the integer optimization problem. However, the scheme for D2D resource allocation should be considered more from practical aspect. The small modulation and simple channel coding are preferable as the devices in massive machine communications are low-complexity [44], so that uplink SCMA system with large users can be resolved in small-scale SCMA structure with same overloading feature, where the SCMA scheme with $J=6$, $K=4$, and $N=2$ is claimed to be the basic regular SCMA system. Besides, the number of D2D pairs is much lower than number of cellular users, which actually reduces the computational complexity. Thus, the analysis in this paper chooses basic search for D2D users' RB allocation after considering the complexity.

When $J_{D}=1$, $\bm{A}$ becomes a $K$-dimensional vector with one non-zero element. Our goal is to find the best column in $\bm{I}_{K}$ to get the best cellular sum-rate while guaranteeing D2D's transmission requirements. For the case of $1<J_{D}\leq K$, let $\lambda\triangleq C_{K}^{J_{D}}$ and $\mu\triangleq J_{D}!$ in the following part. The RB allocation method is summarized in Algorithm~\ref{algorithm2}.
\begin{algorithm}[h]
	\caption{RB allocation algorithm for D2D.}
	\begin{algorithmic}[1]
		\STATE Initialize the set $\Omega=\{1,2,...,K\}$ and list the $\lambda$ subset of $\Omega$ with $J_{D}$ elements as $\{\Omega_{1},...,\Omega_{\lambda}\}$.
		\FOR {$s=1:\lambda$}
		\STATE list the full permutation for elements in $\Omega_{s}$, and let each permutation be a $J_{D}\times1$ vector $\bm{\beta}$. Then, let $\bm{B}=[\bm{\beta}_{1},...,\bm{\beta}_{\mu}]$.
		\FOR {$t=1:\mu$}
		\STATE From $\ell=1,...,J_{D}$, let $k=\bm{B}_{\ell t}$ and set $a_{
			\ell k}=1$ to find $\bm{A}^{(st)}$.
		\IF {Constraint $C1$ is satisfied}
		\STATE Calculate $R^{c(st)}$ with $\bm{A}^{(st)}$.
		\ENDIF
		\ENDFOR
		\ENDFOR
		\RETURN The optimal solution $\bm{A}^{*}$ when the maximum $R^{c(st)}$ is found.
	\end{algorithmic}
	\label{algorithm2}
\end{algorithm}
\subsection{Subproblem of Phase Shifts}
\label{thetaforp1}
This part focuses on the optimization for RIS phase shifts when the variables $\bm{\mathcal{P}}$ and $\bm{A}$ are given. The subproblem $\bm{P7}$ is formulated as
\begin{equation}
\begin{aligned}
&\bm{P7:}\\
&\max_{\bm{\Theta}}\quad \sum_{k=1}^{K}\log\left(1+\frac{\sum_{j\in\xi_{k}}|h_{jk}^{CB}(\bm{\Theta})|^{2}p_{jk}^{c}}{\sum_{\ell=1}^{J_{D}}a_{\ell k}|h_{\ell k}^{DB}(\bm{\Theta})|^{2}p_{\ell}^{d}+N_{0}}\right)\\
&s.t.C1:\frac{\sum_{k=1}^{K}a_{\ell k}|h_{\ell k}^{DD}(\bm{\Theta})|^{2}p_{\ell}^{d}}{\sum_{k=1}^{K}a_{\ell k}(\sum_{j\in\xi_{k}}|h_{jk}^{CD}(\bm{\Theta})|^{2}p_{jk}^{c})+N_{0}}\geq\gamma_{0}^{d},\\
&\quad C6:0<\theta_{m}\leq 2\pi,\quad m=1,\dots,M.\\
\end{aligned}
\label{eq18}
\end{equation}
Notice that the channel conditions with $\bm{\Theta}$ has complex expressions, we decide to handle with these parts. Take $|h_{jk}^{CB}(\bm{\Theta})|^{2}$ for the example. According to \eqref{eq2}, we define
\begin{equation}
\bm{r}_{jk}^{cb}\triangleq diag\{(\bm{g}_{jk}^{cb})^{H}\}\bm{f}_{jk}^{cb},
\label{eq19}
\end{equation}
and
\begin{equation}
\bm{\theta}\triangleq [e^{j\theta_{1}},e^{j\theta_{2}},...,e^{j\theta_{M}}]^{H}.
\label{eq20}
\end{equation}
Then we have
\begin{equation}
\begin{aligned}
|h_{jk}^{cb}&+(\bm{g}_{jk}^{cb})^{H}\bm{\Theta}\bm{f}_{jk}^{cb}|^{2}\\
&=|h_{jk}^{cb}+\bm{\theta}^{H}\bm{r}_{jk}^{cb}|^{2}\triangleq\overline{\bm{\theta}}^{H}\bm{Q}_{jk}^{cb}\overline{\bm{\theta}}+|h_{jk}^{cb}|^{2},\\
\end{aligned}
\label{eq21}
\end{equation}
where $\overline{\bm{\theta}}=[e^{j\theta_{1}},e^{j\theta_{2}},...,e^{j\theta_{M}},1]^{H}$ and
\begin{equation}
\bm{Q}_{jk}^{cb}=\left[ \begin{array}{cc}
\bm{r}_{jk}^{cb}(\bm{r}_{jk}^{cb})^{H} & (h_{jk}^{cb})^{H}\bm{r}_{jk}^{cb} \\
h_{jk}^{cb}(\bm{r}_{jk}^{cb})^{H} & 0
\end{array}\right].
\label{eq22}
\end{equation}
Besides, let $\bm{V}\triangleq\overline{\bm{\theta}}\overline{\bm{\theta}}^{H}$, and the term $\overline{\bm{\theta}}^{H}\bm{Q}_{jk}^{cb}\overline{\bm{\theta}}$ can be replaced as $\overline{\bm{\theta}}^{H}\bm{Q}_{jk}^{cb}\overline{\bm{\theta}}=Tr(\bm{Q}_{jk}^{cb}\bm{V})$. Notice that $\bm{V}$ is a Hermitian matrix with positive semidefinite and $rank(\bm{V})=1$, so the optimization variable $\bm{\Theta}$ has been replaced by $\bm{V}$ to make the problem more tractable. Similarly, we define
\begin{equation}
\begin{aligned}
&Tr(\bm{Q}_{\ell k}^{db}\bm{V})+|h_{\ell k}^{db}|^{2}\triangleq |h_{\ell k}^{DB}(\bm{\Theta})|^{2}\\
&Tr(\bm{Q}_{\ell k}^{dd}\bm{V})+|h_{\ell k}^{dd}|^{2}\triangleq |h_{\ell k}^{DD}(\bm{\Theta})|^{2}\\
&Tr(\bm{Q}_{jk}^{cd}\bm{V})+|h_{jk}^{cd}|^{2}\triangleq |h_{jk}^{CD}(\bm{\Theta})|^{2},
\end{aligned}
\label{eq23}
\end{equation}
then the constraint $C1$ can be rewritten as
\begin{equation}
\begin{aligned}
&\sum_{k=1}^{K}a_{\ell k}[Tr(\bm{Q}_{\ell k}^{dd}\bm{V})+|h_{\ell k}^{dd}|^{2}]p_{\ell}^{d}\\
&\geq\sum_{k=1}^{K}\gamma_{0}^{d}a_{\ell k}\{\sum_{j\in\xi_{k}}[Tr(\bm{Q}_{jk}^{cd}\bm{V})+|h_{jk}^{cd}|^{2}]p_{jk}^{c}\}+\gamma_{0}^{d}N_{0}.
\end{aligned}
\label{eq24}
\end{equation}

Based on the above process, the subproblem $\bm{P7}$ is transformed into the following approximated problem which is shown as
\begin{equation}
\begin{aligned}
\bm{P8:}\quad &\max_{\bm{V}}\quad \sum_{k=1}^{K}[f_{k}(\bm{V})-g_{k}(\bm{V})]\\
&s.t.\quad C14:(24)\\
\quad &C15: \bm{V}_{mm}=1, \quad m=1,\dots,M+1,\\
\quad &C16: \bm{V}\succeq 0, \quad \bm{V}\in\mathbb{H}^{M+1},\\
\quad &C17: rank(\bm{V})=1,
\end{aligned}
\label{eq25}
\end{equation}
where
\begin{equation}
\begin{aligned}
f_{k}(\bm{V})&\triangleq\log\{\sum_{\ell=1}^{J_{D}}a_{\ell k}[Tr(\bm{Q}_{\ell k}^{db}\bm{V})+|h_{\ell k}^{db}|^{2}]p_{\ell}^{d}+N_{0}\\
&+\sum_{j\in\xi_{k}}[Tr(\bm{Q}_{jk}^{cb}\bm{V})+|h_{jk}^{cb}|^{2}]p_{jk}^{c}\},
\end{aligned}
\label{eq26}
\end{equation}
and
\begin{equation}
g_{k}(\bm{V})\triangleq\log\{\sum_{\ell=1}^{J_{D}}a_{\ell k}[Tr(\bm{Q}_{\ell k}^{db}\bm{V})+|h_{\ell k}^{db}|^{2}]p_{\ell}^{d}+N_{0}\}.
\label{eq27}
\end{equation}
The constraint $C16$ denotes that variable $\bm{V}$ is Hermitian and positive semidefinite, and the constraint $C17$ denotes rank-one constraint which is non-convex. Notice that for any $\bm{V}\in \mathbb{H}^{M+1}$, we have $||\bm{V}||_{*}-||\bm{V}||_{2}\geq 0$, where $||\bm{V}||_{*}$ denotes the nuclear norm and $||\bm{V}||_{2}$ denotes the spectral norm. The equality holds if and only if $\bm{V}$ is a rank-one matrix. Thus, we introduce a penalty coefficient $\eta$ and rewrite the optimization problem $\bm{P8}$ as
\begin{equation}
\begin{aligned}
\bm{P9:}&\min_{\bm{V}} \sum_{k=1}^{K}[g_{k}(\bm{V})-f_{k}(\bm{V})]+\eta(||\bm{V}||_{*}-||\bm{V}||_{2})\\
&s.t.\quad C14, C15, C16.\\
\end{aligned}
\label{eq28}
\end{equation}
The problem $\bm{P9}$ is proved to be equivalent to $\bm{P8}$ with $\eta\to\infty$ [45]. However, the objective function is still non-convex, and we utilize successive convex approximation (SCA) for help.

Notice that $g_{k}(\bm{V})$ is a concave function, for the given point $\bm{V}_{0}$, the global upper bound based on the first-order Taylor expansion is shown as
\begin{equation}
g_{k}(\bm{V})\leq\hat{g}_{k}\triangleq g_{k}(\bm{V}_{0})+Tr[(\bigtriangledown g_{k}(\bm{V}_{0}))^{H}(\bm{V}-\bm{V}_{0})].
\label{eq29}
\end{equation}
Similarly, a lower bound for the convex function $||\bm{V}||_{2}$ with point $\bm{V}_{0}$ is given by
\begin{equation}
||\bm{V}||_{2}\geq\hat{\bm{V}}\triangleq ||\bm{V}_{0}||_{2}+Tr[(\bigtriangledown||\bm{V}_{0}||_{2})^{H}(\bm{V}-\bm{V}_{0})].
\label{eq30}
\end{equation}
Then, part of the objective function is approximated in order to formulate the convex optimization problem $\bm{P10}$,
\begin{equation}
\begin{aligned}
\bm{P10:}&\min_{\bm{V}} \sum_{k=1}^{K}[\hat{g}_{k}(\bm{V})-f_{k}(\bm{V})]+\eta(||\bm{V}||_{*}-\hat{\bm{V}})\\
&s.t.\quad C14, C15, C16.\\
\end{aligned}
\label{eq31}
\end{equation}
Notice that problem $\bm{P10}$ can be solved by Matlab CVX. Therefore, our proposed algorithm utilize iterative process to find the stationary point of problem $\bm{P8}$, and then obtain the optimal phase shift matrix $\bm{\Theta}$ from $\bm{V}$. The iterative algorithm is summarized in Algorithm~\ref{algorithm3}.
\begin{algorithm}[h]
	\caption{Iterative RIS phase shift allocation algorithm.}
	\begin{algorithmic}[1]
		\STATE Initialize random value for $\bm{V}^{(0)}$, $\eta$, and penalty gain $pg_{\eta}$. Set the maximum iterative number $T_{3}$ and $T_{4}$.
		\STATE Set $I_{3}=1$.
		\WHILE {$I_{3}\leq T_{3}$}
		\STATE Set $I_{4}=1$.
		\WHILE {$I_{4}\leq T_{4}$}
		\STATE Solve problem $\bm{P10}$ by $\bm{V}^{(I_{4}-1)}$ to find optimal solution $\bm{V}^{*(I_{4})}$.
		\STATE Update $\bm{P10}$ with new $\bm{V}^{(I_{4})}$.
		\STATE $I_{4}=I_{4}+1$.
		\ENDWHILE     		
		\STATE Update $\eta=pg_{\eta}\cdot\eta$.
		\STATE $I_{3}=I_{3}+1$.
		\ENDWHILE
		\RETURN The optimal solution $\bm{V}^{*}$ for $\bm{P8}$.
		\STATE Calculate the eigenvalue and eigenvector of $\bm{V}^{*}$ to obtain the optimal phase shifts $\bm{\Theta}^{*}$ for $\bm{P7}$.
	\end{algorithmic}
	\label{algorithm3}
\end{algorithm}

\subsection{The Optimality of Proposed Algorithm}
\label{overall}
\begin{algorithm}
	\caption{Overall iterative allocation algorithm of $\bm{P1}$.}
	\begin{algorithmic}[1]
		\STATE Initialize set $\{\bm{\mathcal{P}}^{(0)},\bm{A}^{(0)},\bm{\Theta}^{(0)}\}$. Set the maximum iterative number $\tilde{N}$.
		\STATE Set $n=0$.
		\WHILE {$n\leq \tilde{N}$}
		\STATE For given $\bm{\mathcal{P}}^{(n)}$, $\bm{A}^{(n)}$, $\bm{\Theta}^{(n)}$, solve the problem $\bm{P2}$ by Algorithm~\ref{algorithm1} to obtain $\bm{\mathcal{P}}^{(n+1)}$.    		
		\STATE For given $\bm{\mathcal{P}}^{(n+1)}$, $\bm{\Theta}^{(n)}$, solve the problem $\bm{P6}$ by Algorithm~\ref{algorithm2} to obtain $\bm{A}^{(n+1)}$.
		\STATE For given $\bm{\mathcal{P}}^{(n+1)}$, $\bm{A}^{(n+1)}$, and $\bm{\Theta}^{(n)}$, solve the problem $\bm{P7}$ by Algorithm~\ref{algorithm3} to obtain $\bm{\Theta}^{(n+1)}$.
		\STATE $n=n+1$.
		\ENDWHILE
		\RETURN The final solution set $\{\bm{\mathcal{P}}^{*},\bm{A}^{*},\bm{\Theta}^{*}\}$.
		\STATE	Calculate the optimal value for cellular sum-rate $R^{c}$.
	\end{algorithmic}
	\label{algorithm4}
\end{algorithm}

Based on the analysis for three subproblems, the overall algorithm for solving optimization problem $\bm{P1}$ is summarized in Algorithm~\ref{algorithm4}.

We have analyzed that the original problem $\bm{P1}$ is non-convex and NP-hard. Then, the three subproblems are iteratively solved based on BCD method [12], [45]. Specifically, some non-convex constraints are replaced based on CUB approximation in Algorithm~\ref{algorithm1}, so the solution of problem $\bm{P2}$ is suboptimal. Thereafter, the search method in Algorithm~\ref{algorithm2} can find the optimal solution of problem $\bm{P6}$. Besides, in Algorithm~\ref{algorithm3} we utilize SCA to handle the non-convex constraints, and the solution of problem $\bm{P8}$ is still suboptimal. As the algorithm in Algorithm~\ref{algorithm4} is guaranteed to converge to a stationary point, the final solution $\{\bm{\mathcal{P}}^{*},\bm{A}^{*},\bm{\Theta}^{*}\}$ after all iterations should be the suboptimal solution of original problem $\bm{P1}$. 

\subsection{Convergence and Complexity Analysis}
\label{cca}
\subsubsection{Convergence}
In Algorithm~\ref{algorithm1}, we need $T_{1}$ iterations for problem $\bm{P5}$ and $T_{2}$ iterations for problem $\bm{P4}$. After the $i$-th iteration process of the $T_{1}$ iterations, the objective value $z$ can be denoted as
\begin{equation}
z^{(i)}=z\left(\bm{\mathcal{P}}^{(i)},\bm{\mathcal{Q}}^{(i)}\right).
\label{eq32}
\end{equation}
Notice that when $\bm{\alpha}$ is updated by $\alpha_{\ell k}=\frac{p_{\ell}^{d}}{Q_{k}}$, the objective value $z$ keeps same. Then
\begin{equation}
z^{(i)}=z\left(\bm{\mathcal{P}}^{(i)},\bm{\mathcal{Q}}^{(i)},\bm{\alpha}^{(i)}\right)=z\left(\bm{\mathcal{P}}^{(i)},\bm{\mathcal{Q}}^{(i)},\bm{\alpha}^{(i+1)}\right).
\label{eq33}
\end{equation}
By using $\bm{\alpha}^{(i+1)}$ for the next iteration, when we obtain the optimal solution, we must have
\begin{equation}
z\left(\bm{\mathcal{P}}^{(i+1)},\bm{\mathcal{Q}}^{(i+1)},\bm{\alpha}^{(i+1)}\right)\leq z\left(\bm{\mathcal{P}}^{(i)},\bm{\mathcal{Q}}^{(i)},\bm{\alpha}^{(i+1)}\right),
\label{eq34}
\end{equation}
thus, we have
\begin{equation}
z^{(i+1)}=z\left(\bm{\mathcal{P}}^{(i+1)},\bm{\mathcal{Q}}^{(i+1)},\bm{\alpha}^{(i+1)}\right)\leq z^{(i)}.
\label{eq35}
\end{equation}
According to the limitation in problem $\bm{P5}$, the objective value has the lower bound. When the value of iterations is large enough, this optimal value is guaranteed to converge. Similarly, the iterations for problem $\bm{P4}$ can also be proved to converge to a stationary point based on above analysis. Then, the Algorithm~\ref{algorithm1} is convergent.

Next the convergence of Algorithm~\ref{algorithm3} is analyzed. During the inner $T_{4}$ iterations, the problem $\bm{P10}$ is always a convex optimization problem and the objective function is monotonically non-increasing. Then, the inner process is guaranteed to converge. For the outer iterations, when the penalty coefficient increases to large enough, the objective values can converge to a stationary point.

For the overall algorithm in Algorithm~\ref{algorithm4}, we define that
\begin{equation}
R^{c(n)}\triangleq R^{c}\left(\bm{\mathcal{P}}^{(n)},\bm{A}^{(n)},\bm{\Theta}^{(n)}\right).
\label{eq36}
\end{equation}
As the problem $\bm{P2}$, $\bm{P6}$, and $\bm{P7}$ are solved during the $n$-th iteration, we must have
\begin{equation}
R^{c}\left(\bm{\mathcal{P}}^{(n)},\bm{A}^{(n)},\bm{\Theta}^{(n)}\right)\leq R^{c}\left(\bm{\mathcal{P}}^{(n+1)},\bm{A}^{(n+1)},\bm{\Theta}^{(n+1)}\right).
\label{37}
\end{equation}
Notice that the objective function has the upper bound because of power limitation, the proposed algorithm in Algorithm~\ref{algorithm4} is proved to converge to a stationary point.

\subsubsection{Complexity}
In Algorithm~\ref{algorithm1}, the computational complexity is $\bm{\mathcal{O}}(T_{1}(w+1)^{3.5}+T_{2}(w)^{3.5})$, where $w\triangleq NJ+J_{D}+K$. The computational complexity of Algorithm~\ref{algorithm2} is $\bm{\mathcal{O}}(\lambda\mu)$. For Algorithm~\ref{algorithm3}, the computational complexity of solving problem $\bm{P10}$ is $\bm{\mathcal{O}}\left((M+2)^{4.5}\right)$, so the computational complexity of applying Algorithm~\ref{algorithm3} is $\bm{\mathcal{O}}\left(T_{3}T_{4}(M+2)^{4.5}\right)$. Thus, for the overall proposed algorithm in Algorithm~\ref{algorithm4}, the computational complexity is $\bm{\mathcal{O}}\left(\tilde{N}(T_{1}(w+1)^{3.5}+T_{2}(w)^{3.5}+\lambda\mu+T_{3}T_{4}(M+2)^{4.5})\right)$.

\section{Simulation Results}
\label{Simulation}
In this section, we perform simulations to demonstrate the performance of our proposed resource allocation algorithm. Our simulation model utilizes a single-cell uplink SCMA system. Notice that the small modulation and simple channel coding are preferable as the devices in massive machine communications are low-complexity [44], the uplink SCMA system with large users and massive RISs can be resolved in small-scale SCMA structure and RISs with same overloading feature. Thus, our scheme utilizes $J=6$ and $K=4$ with a small RIS. For the location scenario, the BS and RIS are located at coordinates (0,0,15) meters and (300, 0, 15) meters, respectively. The cellular users are randomly and uniformly placed in a circle centered at (320, 0, 0) meters with radius $10$ m, while the D2D pairs are randomly and uniformly placed in a circle centered at (500, 0, 0) meters with radius $10$ m. In each D2D pair, the distance between $DT$ and $DR$ is uniformly and independently distributed from $[1,5]$ meters. The channel models for both cellular communications and D2D communications are generated with a normalized Rayleigh fading component as small-scale fading and a distance-dependent path loss model as big-scale fading. For the SINR requirements of D2D communications, we set the minimum rate $R_{0}^{d}$ instead of $\gamma_{0}^{d}$ as $\gamma_{0}^{d}=2^{R_{0}^{d}}-1$. Besides, other simulation parameters are summarized in Table~\ref{table1}.

\begin{table}
	\centering
	\caption{Simulation parameters.}
	\begin{tabular}{|c|c|}
		\hline
		\scriptsize No. CU, $J$ & \scriptsize 6\\
		\hline
		\scriptsize No. subcarriers, $K$ & \scriptsize 4\\
		\hline
		\scriptsize No. reflecting elements, $M$ & \scriptsize 4\\
		\hline
		\scriptsize Noise power, $N_{0}$ & \scriptsize -174dBm/Hz\\
		\hline
		\scriptsize Power limitation of CU, $P_{0}^{c}$ & \scriptsize 30dBm\\
		\hline
		\scriptsize Power limitation of DU, $P_{0}^{d}$ & \scriptsize 30dBm\\
		\hline
		\scriptsize Rate limitation of DU, $R_{0}^{d}$ & \scriptsize 30bps\\
		\hline
		\scriptsize Path loss Model of D2D & \scriptsize 40log10($d_D$[km])+148\\
		\hline
		\scriptsize Path loss Model of CU & \scriptsize 37.6log10($d_C$[km])+128.1\\
		\hline
	\end{tabular}
	\label{table1}
\end{table}

In order to evaluate the performance of our proposed algorithm, we consider some different benchmark methods. For RPS scheme, the power variables and RB allocation variables are optimized with our proposed methods while the phase shift matrix $\bm{\Theta}$ are randomly allocated, as each phase shift is uniformly and independently generated from $(0,2\pi]$. For RPO scheme, each user's power is randomly allocated under constraints while other variables are optimized. Besides, D2D users randomly select RB in RRB scheme with optimized power and RIS phase shifts. In the last scheme, the system performance is simulated without RIS.

\subsection{Algorithm Performance versus D2D Pair}
\label{per1}
\begin{figure}[h]
	\centering  
	\includegraphics[width=0.9\linewidth]{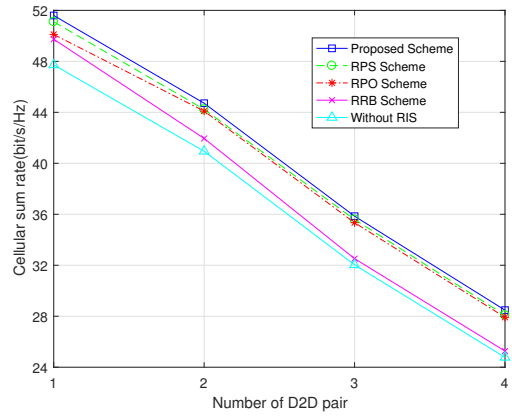}  
	\caption{Cellular sum-rate versus number of D2D pairs.} 
	\label{fig2}
\end{figure}

The cellular sum-rate performance versus different number of D2D pairs is shown in Fig.~\ref{fig2}. The parameters of $P_{0}^{c}$, $P_{0}^{d}$, and $R_{0}^{d}$ are set as in Table~\ref{table1}. As the number of D2D pairs $J_{D}$ is not more than $K$, Fig.~\ref{fig2} gives all four situations. Firstly, the figure shows that the increase of D2D pairs leads to the reduce of cellular sum-rate. This is obviously because the mutual interference between cellular users and D2D users becomes stronger with more D2D communications. We can observe that every additional D2D pair makes the performance of cellular sum-rate reduce about $8$ bit/s/Hz.

Besides, the comparison with different schemes shows that our proposed algorithm outperforms other four schemes. The proposed scheme significantly improves the sum-rate performance compared with RRB scheme and the scheme without RIS, which proves that RIS-aided networks can bring considerable gain for SCMA cellular communications, and the lack for the optimization of RB allocation problem will bring worst influence to cellular sum-rate performance. Notice that the gap between proposed scheme and RPS scheme are always about $1$ bit/s/Hz with different D2D pairs. The probable reason is that the RIS is not large enough and the reflecting elements are limited, which makes the limited optimization of phase shift. Thereafter, the gap between proposed scheme and RPO scheme also becomes small as D2D pairs increase. That is because more mutual interference is considered when more users' power needs to be optimized, and some power variables have to be limited optimized.

\begin{figure}[h]
	\centering  
	\includegraphics[width=0.9\linewidth]{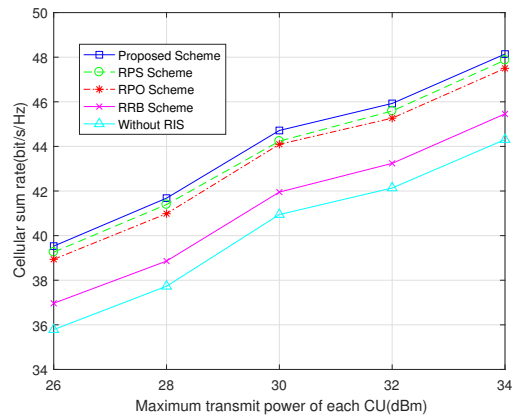}  
	\caption{Cellular sum-rate versus CU power limitation.} 
	\label{fig3}
\end{figure}

\subsection{Algorithm Performance versus Cellular Power Limitation}
\label{per2}
We compare the performance of the proposed algorithm with other four schemes versus different $CU$ power limitation, $P_{0}^{c}$, in this part. The scenario of D2D pairs is set as $J_D=2$. From Fig.~\ref{fig3} we can find that as the maximum power improves, the performance of cellular sum-rate also improves smoothly. The reason is that more cellular users can be allocated with high value power while the requirements of D2D communications are guaranteed. Then, the proposed algorithm still presents best performance, as it always brings an improvement of $4$ bit/s/Hz compared with the scheme without RIS. Besides, the performance is also insignificant compared with RPS scheme as the limited elements. The gap between proposed scheme and RPO scheme becomes larger, while the RRB scheme still makes worst performance.

\begin{figure}[h]
	\centering  
	\includegraphics[width=0.9\linewidth]{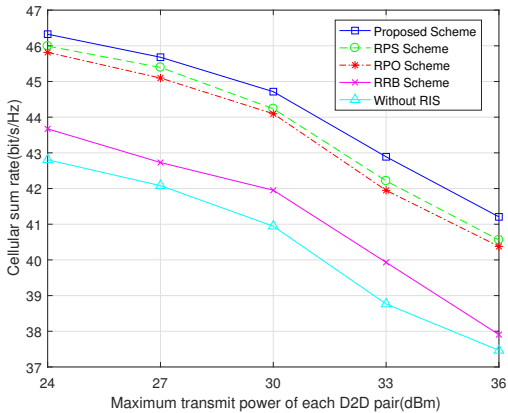}  
	\caption{Cellular sum-rate versus D2D power limitation.} 
	\label{fig4}
\end{figure}
\begin{figure}[h]
	\centering  
	\includegraphics[width=0.9\linewidth]{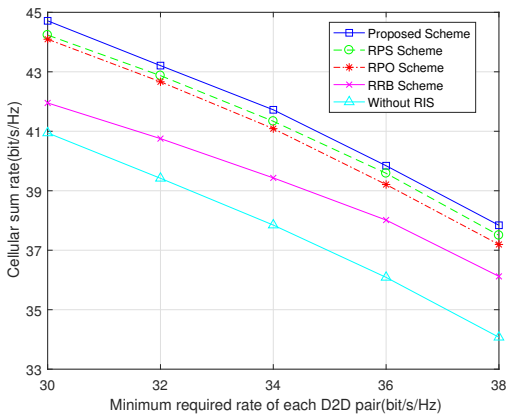}  
	\caption{Cellular sum-rate versus D2D rate limitation.} 
	\label{fig5}
\end{figure}

\subsection{Algorithm Performance versus D2D Power Limitation}
\label{per3}
Fig.~\ref{fig4} focuses on the cellular sum-rate performance versus different D2D power limitation, $P_{0}^{d}$. We still set $J_D=2$ for D2D pairs and keep other basic parameters invariant. Different from the trend in Fig.~\ref{fig3}, the cellular sum-rate reduces smoothly when the D2D power limitation increases. It is easy to understand as each D2D pair with higher value power brings stronger interference to cellular communications. On the other hand, the proposed algorithm still shows more significant gain compared with other schemes. The gap between proposed scheme and four other schemes basically keeps the same as shown in Fig.~\ref{fig3}, and the similar reason is omitted here.

\subsection{Algorithm Performance versus D2D Rate Limitation}
\label{per4}
In this part, we operate simulation to observe the performance of cellular sum-rate versus different D2D rate limitation, $R_{0}^{d}$. The scenario of other parameters is the same as above parts. Fig.~\ref{fig5} also shows the monotonic decreasing trend, as cellular communications have to make sacrifice to improve the qualities of D2D communications. Besides, the proposed algorithm with RIS-aided network still keeps outstanding performance, which proves that RIS brings prominent improvement for SCMA cellular network to mitigate the mutual interference as much as possible.

We can also notice that the gap between RRB scheme and proposed scheme becomes smaller with higher D2D rate limitation. That is because when the requirements of D2D communications improve, there are fewer options for each D2D pair to select corresponding RB. Thus, the random distribution becomes closer to the optimal distribution.

\section{Conclusion}
\label{Conclusion}
In this paper, we utilize RIS technologies and SCMA schemes to improve the cellular communication quality of our hybrid network with D2D users. To improve the performance of cellular sum-rate, we focus on the joint resource allocation problem including user's power, phase shifts at the RIS, and the RB allocation of D2D pairs. The problem with coupled variables is decoupled into three subproblems and we propose corresponding methods to solve them. Simulation results prove that our proposed algorithm is appropriate and efficient to solve the optimization problem. Besides, reinforcement learning (RL) algorithms are becoming promising methods for solving optimization problems. In our further research with RIS-aided uplink SCMA system coexisting with D2D, we will aim to explore the potential of this approach.

.

\EOD

\end{document}